\documentclass[submission,copyright,creativecommons]{eptcs}

\providecommand{\event}{GASCom} 

\usepackage{iftex}

\ifpdf
  \usepackage{underscore}         
  \usepackage[T1]{fontenc}        
\else
  \usepackage{breakurl}           
\fi

\usepackage{graphicx}
\usepackage{nameref}
\usepackage{amsthm}
\usepackage{amsmath}
\usepackage{amsfonts} 
\usepackage{enumerate}
\usepackage{stmaryrd}
\usepackage{xassoccnt}
\usepackage{hyperref}
\usepackage{cleveref}
\usepackage{mathrsfs}
\usepackage[T1]{fontenc}
\usepackage[english]{babel}
\usepackage{thmtools}
\usepackage{thm-restate}

\newtheorem{theorem}{Theorem}[section]
\crefname{theorem}{theorem}{theorems}
\newtheorem{proposition}{Proposition}[section]
\crefname{proposition}{proposition}{propositions}
\newtheorem{lemma}{Lemma}[section]
\crefname{lemma}{lemma}{lemmas}

\theoremstyle{definition}

\crefname{remark}{remark}{remarks}

\crefname{example}{example}{examples}

\DeclareCoupledCountersGroup{theorems}
\DeclareCoupledCounters[name=theorems]{theorem,proposition,lemma,remark,example}

\title{Morphic Sequences: Complexity and Decidability}
\author{Raphael Henry
\institute{I2M\\ Marseille, France}
\email{raphael.henry@univ-amu.fr}
}
\date{\sc Extended Abstract \\
 \\
21/05/2024}

\newcommand{\titlerunning}{Morphic Sequences: Complexity and Decidability}
\newcommand{\authorrunning}{R. Henry}

\hypersetup{
  bookmarksnumbered,
  pdftitle    = {\titlerunning},
  pdfauthor   = {\authorrunning},
  pdfsubject  = {EPTCS},               
}

\begin{document}
\maketitle

\begin{abstract}

In this work we recall Pansiot's result on the complexity of pure morphic sequences and we use the tools developed by Devyatov for morphic sequences to prove the decidability of the complexity class of pure morphic sequences.
\end{abstract}

\section{Introduction}

In symbolic dynamics, a natural way to generate right infinite words (indexed by $\mathbb{N} = \{0,1,2...\}$) on a finite alphabet $A$ is to iterate a morphism $\varphi : A^* \rightarrow A^*$ on a letter $a$, a process which converges to a fixed point of $\varphi$. We call such a word a \textbf{pure morphic sequence} and we denote it by $\varphi^\infty(a)$. More generally, applying a coding $\psi$, that is a letter-to-letter morphism, to a pure morphic word gives a \textbf{morphic sequence} denoted by $\psi(\varphi^\infty(a))$.

A major tool of symbolic dynamics is the \textbf{factor complexity}: the function $P_\alpha : \mathbb{N} \rightarrow \mathbb{N}$ counting the number of rows (factors) of length $n$ appearing in the sequence $\alpha$. An important result linking complexity and the structure of sequences is the following:

\begin{theorem}[Morse-Hedlund, 1938]\label{MorseHedlund}
    A sequence $\alpha$ is ultimately periodic if and only if $P_\alpha(n) \leq n$ for some $n \in \mathbb{N}^*$ if and only if $P_\alpha(n)$ is bounded.
\end{theorem}

In this work we study the characterization of the complexity of pure morphic sequences. For example, with D0L-systems, Ehrenfeucht, Lee et Rozenberg showed in 1975 that the complexity of pure morphic sequences is $\mathcal{O}(n^2)$. Other lower and upper bounds were obtained in particular cases until Pansiot gave the complete classification in \cite{Pansiot84} using criteria on the morphism:

\begin{restatable}[J.J. Pansiot, 1984]{theorem}{pansiot}\label{Pansiot}
	The complexity of pure morphic sequences belongs to one of the five classes:
	\begin{equation*}
		\Theta(1)~,~\Theta(n)~,~\Theta(n \log \log n)~,~\Theta(n \log n)~,~\Theta(n^2).
	\end{equation*}
\end{restatable}

Applying a coding will either permute letters or merge some of them, which can only decrease the complexity. Doing so, new complexity classes appear:

\begin{proposition}[J.J. Pansiot, 1985]
    For every $k \in \mathbb{N}^*$, there exists a morphic sequence $\alpha$ such that $P_\alpha(n) = \Theta(n^{1+1/k})$.
\end{proposition}

This result is stated in \cite{Pansiot85}, and the example of a pure morphic sequence of complexity $\Theta(n^{1+1/k})$ is detailed in \cite{CANT}:
\begin{itemize}
	\item $A=\{a, b_0, b_1, ... b_k\}$
	\item $\varphi(a)=a b_k$, $\varphi(b_0)=b_0$ and $\varphi(b_i)=b_i b_{i-1}$ for $i \in \llbracket 1,k \rrbracket$
	\item $\psi(a)=0$, $\psi(b_i)=0$ for $i \in \llbracket 0,k-1 \rrbracket$ and $\psi(b_k)=1$
\end{itemize}
~\\
In \cite{Devyatov}, Devyatov shows that they are the only classes between $\Theta(n \log n)$ et $\Theta(n^2)$:

\begin{theorem}[R. Devyatov, 2015]\label{Devyatov}
    The complexity of morphic sequences is either:
    \begin{equation*}
    	\Theta(n^{1+1/k}) \textrm{ for some } k \in \mathbb{N}^*~~~\textrm{or}~~~\mathcal{O}(n \log n).
	\end{equation*}
\end{theorem}

\section{Result}

In \cite{Pansiot85}, Pansiot mentions the following decidability problem:
\begin{align*}
   \textbf{PMClass:~~~~~} & \textrm{\textbf{Input:} A pure morphic sequence } \alpha=\varphi^\infty(a) \\
     & \textrm{\textbf{Question:} What is the complexity class of } \alpha \textrm{ ?}
\end{align*}

\Cref{Pansiot} states that there are five possible answers and its proof exhibits criteria for each complexity class, which we will formulate in an algorithm. By deciding every criterion, we prove the following result:

\begin{theorem}
    PMClass is decidable.
\end{theorem}

To achieve that, we use the detailed proof of \Cref{Pansiot} in \cite{CANT} and a decidability result from Pansiot \cite{Pansiot86} and Harju-Linna \cite{Linna}, and we adapt some parts of Devyatov's proof.

\section{Sketch of the proof}

\subsection{Growth of morphism}

If $a \in A$ is a letter, the \textbf{growth rate} of $a$ is the function associated to the asymptotic behaviour of $\lvert \varphi^k(a) \rvert$ when $k$ tends to $\infty$. The following theorem stated in \cite{OGSalomaa} gives its precise form:

\begin{theorem}[A. Salomaa, M. Soittola]\label{Salomaa}
    For each morphism $\varphi : A^* \rightarrow A^*$ and each letter $a\in A$, there exist $(\beta,\alpha)\in(\mathbb{R}_{\geq1}\times \mathbb{N})\cup \{(0,0)\}$ such that 
    \begin{equation*}
    \lvert \varphi^k(a) \rvert=\Theta(k^\alpha \beta^k).
    \end{equation*}
\end{theorem}

We say a letter is \textbf{bounded} if its growth rate is bounded ($(\beta_a, \alpha_a) \in \{(0,0),(1,0)\}$), and \textbf{growing} in the other case. We denote by $B$ the set of bounded letters and $C$ the set of growing letters. If every letter is growing, $\varphi$ is said to be \textbf{growing}. The case $(\beta_a, \alpha_a) = (0,0)$ means that $\varphi$ erases the letter $a$ ($\varphi(a)=\varepsilon$ the empty word).

We say $\varphi$ is \textbf{quasi-uniform} if every letter has the same rate of the form $\beta^k$ with $\beta>1$.We say $\varphi$ is \textbf{polynomially divergent} if every letter $a$ has a rate of the form $k^{\alpha_a}\beta^k$ with $\beta>1$, and at least one of the $\alpha_a$ is not 0. We say $\varphi$ is \textbf{exponentially divergent} if there are two letters $a$ and $b$ of rate $k^{\alpha_a}\beta_a^k$ et $k^{\alpha_b}\beta_b^k$ with $1<\beta_a<\beta_b$ and $\beta_c>1$ for all $c \in A$.

These three classes of morphisms are mutually exclusive, and a morphism is growing if and only if it belongs to one of them.

\subsection{Pansiot criteria}

Given a finite alphaet $A$, a morphism $\varphi : A^* \rightarrow A^*$ and a pure morphic sequence, the proof of \Cref{Pansiot} gives the criteria to determine its complexity class.

In particular, the case where $\varphi$ is not growing and the factors of $\alpha$ in $B^*$ have bounded length boils down to computing the complexity of another pure morphic sequence:

\begin{proposition}
If $\varphi$ is not growing and the factors of $\alpha$ in $B^*$ have bounded length, one can explicitly compute an alphabet $\Sigma$, a growing morphism $\sigma : \Sigma^* \rightarrow \Sigma^*$, a letter $b \in \Sigma$ and a non-erasing morphism $\psi : \Sigma^* \rightarrow A^*$ such that 
\begin{equation*}
	\alpha=\psi(\sigma^\infty(b)).
\end{equation*}
Moreover $\alpha$ and $\sigma^\infty(b)$ are in the same complexity class.
\end{proposition}

We formulate the classification with the following algorithm:
\begin{align*}
    & \textrm{\textbf{PMClass}(} \alpha=\varphi^\infty(a) \textrm{):} \\
    & \textrm{if } \alpha \textrm{ is eventually periodic:} \\
    & ~~~~~~~~ \textrm{return } "\Theta(1)" \\
    & \textrm{if } \varphi \textrm{ is growing:} \\
    & ~~~~~~~~ \textrm{if } \varphi \textrm{ is quasi-uniform:} \\
    & ~~~~~~~~~~~~~~~~ \textrm{return } "\Theta(n)" \\
    & ~~~~~~~~ \textrm{if } \varphi \textrm{ is polynomially divergent:} \\
    & ~~~~~~~~~~~~~~~~ \textrm{return } "\Theta(n \log\log n)" \\
    & ~~~~~~~~ \textrm{if } \varphi \textrm{ is exponentially divergent:} \\
    & ~~~~~~~~~~~~~~~~ \textrm{return } "\Theta(n \log n)" \\
    & \textrm{else:} \\
    & ~~~~~~~~ \textrm{if the factors of } \alpha \textrm{ in } B^* \textrm{ have bounded length:} \\
    & ~~~~~~~~~~~~~~~~ \textrm{compute } \Sigma, \sigma, \psi \textrm{ et } b \textrm{ such that } \alpha=\psi(\sigma^\infty(b)) \\
    & ~~~~~~~~~~~~~~~~ \textrm{return } PMClass(\sigma^\infty(b)) \\
    & ~~~~~~~~ \textrm{else:} \\
    & ~~~~~~~~~~~~~~~~ \textrm{return } "\Theta(n^2)"    
\end{align*}

When $\varphi$ is not growing and the factors of $\alpha$ in $B^*$ have bounded length, the algorithm is recursive but the new morphism $\sigma$ is growing so there is only one more iteration. Each complexity class is non-empty, here are examples for each one:
\newline
\newline
• $\Theta(1)$: with $\varphi : a \mapsto ab, b \mapsto c, c \mapsto b$, $\varphi^\infty(a)=abcbcbcbc...$ is eventually periodic.
\newline
\newline
• $\Theta(n)$: with the Thue-Morse morphism $\varphi : a \mapsto ab, b \mapsto ba$, $\lvert \varphi^k(a) \rvert = \lvert \varphi^k(b) \rvert = 2^k$, $\varphi$ is quasi-uniform and $\varphi^\infty(a)=abbabaabbaababba...$.
\newline
\newline
• $\Theta(n \log \log n)$: with $\varphi : a \mapsto aba, b \mapsto bb$, $\lvert \varphi^k(b) \rvert = 2^k$ and $\lvert \varphi^k(a) \rvert = k2^{k-1}+2^k$, $\varphi$ is polynomially divergent and $\varphi^\infty(a)=ababbababbbbababbaba...$.
\newline
\newline
• $\Theta(n \log n)$ : with $\varphi : a \mapsto abc, b \mapsto bb, c \mapsto ccc$, $\lvert \varphi^k(b) \rvert = 2^k$, $\lvert \varphi^k(c) \rvert = 3^k$ and $\lvert \varphi^k(a) \rvert = 2^k + (3^k-1)/2$, $\varphi$ is exponentially divergent and $\varphi^\infty(a)=abcbbcccbbbbccccccccc...$.
\newline
\newline
• $\Theta(n^2)$: with $\varphi : a \mapsto ab, b \mapsto bc, c \mapsto c$, $B = \{c\}$,
 $\varphi^\infty(a)=abc^1bc^2bc^3...$ and for all $i$ $c^i \sqsubset \alpha$.
\newline
\newline
• $\Theta(n)$ in two iterations: with $\varphi : a \mapsto acb, b \mapsto bca, c \mapsto c$, $B = \{c\}$ but the only factors of $\alpha$ in $B^*$ are $\varepsilon$ and $c$. Actually $\varphi^\infty(a) = \psi(\sigma^\infty(a))$ with $\sigma$ being the Thue-Morse morphism on $\{a,b\}$ and $\psi : a \mapsto ac, b \mapsto bc$.

\section{Decidability}

We prove that each condition of the algorithm is decidable.

\subsection{$\alpha$ is eventually periodic}

Pansiot and Harju-Linna proved simultaneously in \cite{Pansiot86} and \cite{Linna} that 
the eventual periodicity can be reduced to properties on factors and prove the decidability:

\begin{theorem}
    The eventual periodicity of pure morphic sequences is decidable.
\end{theorem}

\subsection{$\varphi$ is growing}

Deciding if $\varphi$ is growing, quasi-uniform, polynomially or exponentially divergent can be done by algorithmically computing and comparing eigenvalues of integer matrices.

\subsection{$\varphi$ is not growing}

In order to decide if the factors of $\alpha$ in $B^*$ have bounded length, we prove that it is equivalent to a constructive property on the images of letters. Let us remark that this property appears with a non-constructive form in \cite{Pansiot84} (proof of Theorem 4.1) and in \cite{Durand} (Lemma 3.15). To achieve that we use the notion of $k$-blocks developed by Devyatov in \cite{Devyatov}:

An occurence of $\alpha$ is a factor associated to the position of its letters in $\alpha$, denoted $\alpha_{i...j}$. A \textbf{1-block} is an occurence of $\alpha$ in $B^*$ surrounded by two growing letters that we call the \textbf{left border} and the \textbf{right border}. Then $\alpha$ can be split into an alternation of (possibly empty) 1-blocks and growing letters.

If $u$ is a 1-block, then $\varphi(u)$ is an occurence of $\alpha$ containing only bounded letters so it is contained in a unique 1-block that we call the \textbf{descendant} of $u$ and we denote it by $Dc_1(u)$.

Also if there exists a 1-block $v$ such that $Dc_1(v)=u$, then $v$ is unique, we call it the \textbf{ancestor} of $u$ and we denote it by $Dc_1^{-1}(u)$. If a 1-block has no ancester, which is equivalent to the fact that the 1-block and its borders are contained in the image of a letter under $\varphi$, we say it is an \textbf{origin}.

An \textbf{evolution} of 1-blocks is a sequence $\mathscr E$ of 1-blocks such that ${\mathscr E}_0$ is an origin and, for every integer $l$, ${\mathscr E}_l=Dc_1^l({\mathscr E}_0)$. In particular every 1-block belongs to an evolution of 1-blocks.

For every word $u$ containing a growing letter, $LB(u)$ (resp. $RB(u)$) denotes the longest prefix (resp. suffix) of $u$ in $B^*$, and $LC(u)$ (resp. $RC(u)$) denotes the first (resp. last) growing letter in $u$. With these notations we get a first idea of the structure of 1-blocks.

\begin{lemma}\label{blocB}
    Let $\alpha=\varphi^\infty(a)$ a pure morphic sequence, ${\mathscr E}_l$ a 1-block of index $l$ in its evolution $\mathscr E$ and $\alpha_i, \alpha_j$ the borders of ${\mathscr E}_0$. Then
    \begin{equation*}
        {\mathscr E}_l = RB(\varphi^l(\alpha_i))~\varphi^l({\mathscr E}_0)~LB(\varphi^l(\alpha_j)).
    \end{equation*}
\end{lemma}

For every growing letter $c$ , we also define the following objects:
\begin{equation*}
    \begin{aligned}
        LE(c)=\varphi(LB(\varphi(c))) ~~~~~~~~~~~~~~&~~~~~~~~~~~ RE(c)=\varphi(RB(\varphi(c)))) \\
        LK(c)=LB(\varphi(LC(\varphi(c)))) ~~~~~~&~~~~~~~~~~~ RK(c)=RB(\varphi(RC(\varphi(c)))) \\
        LP(c)=\varphi(LK(c)) ~~~~~~~~~~~~~~~~~~~&~~~~~~~~~~~ RP(c)=\varphi(RK(c))
    \end{aligned}
\end{equation*}

In order to refine the structure, we replace $\varphi$ by a large enough power of itself, which does not modify $\alpha$ nor the properties of the morphism, so that the morphism is \textbf{strongly 1-periodic}. We must note that this power can be bounded using the size of $A$, which makes this process effective.

\begin{lemma}\label{périodeB}
    Let $\varphi$ a strongly 1-periodic morphism. Then for every growing letter $c$ and for every $l \geq 2$
    \begin{equation*}
        LB(\varphi^{l}(c)) = LE(c) ~ LP(c)^{l-2} ~ LK(c)
    \end{equation*}
    \begin{equation*}
        RB(\varphi^l(c)) = RK(c) ~ RP(c)^{l-2} ~ RE(c)
    \end{equation*}
\end{lemma}

These two lemmas lead us to state an equivalent condition which is clearly decidable:

\begin{proposition}
    Let $\varphi$ a strongly 1-periodic morphism and $\alpha=\varphi^\infty(a)$ a non-eventually-periodic pure morphic sequence. Then the factors of $\alpha$ in $B^*$ have bounded length if and only if
    \begin{equation*}
    	\textrm{for every } c \in C,~LP(c)=RP(c)=\varepsilon.
    \end{equation*}
\end{proposition}

\bibliographystyle{eptcs}
\bibliography{generic}
\end{document}